\documentclass[preprint,tightenlines,floatfix,nofootinbib,superscriptaddress,fleqn]{revtex4}
\usepackage{bm}
\usepackage{dcolumn}
\usepackage{amsmath}
\usepackage{amsfonts}
\usepackage{amssymb}
\usepackage{amscd}
\usepackage{amsthm}
\usepackage{graphicx}
\begin{document}
\preprint{PNU-NTG-15/2004}
\title{Magnetic moments of the pentaquarks\footnote{A talk given at
International Workshop on PENTAQUARK04, SPring-8, Hyogo, Japan,
20-23 Jul 2004}} 
\author{Hyun-Chul Kim}
\email{hchkim@pusan.ac.kr}
\affiliation{Department of Physics and  Nuclear Physics \& Radiation
Technology Institute (NuRI), Pusan National University, 609-735
Busan, Republic of Korea}

\author{Ghil-Seok Yang}
\email{hchkim@pusan.ac.kr}
\affiliation{Department of Physics and  Nuclear Physics \& Radiation
Technology Institute (NuRI), Pusan National University, 609-735
Busan, Republic of Korea}
\author{Micha{\l} Prasza{\l}owicz} 
\email{michal@th.if.uj.edu.pl}
\affiliation{M. Smoluchowski Institute of Physics, Jagellonian University, ul.
Reymonta 4, 30-059 Krak{\'o}w, Poland}
\author{Klaus Goeke}
\email{Klaus.Goeke@tp2.ruhr-uni-bochum.de}
\affiliation{Institut f\"ur Theoretische Physik II, Ruhr-Universit\" at Bochum,
D--44780 Bochum, Germany}
\date{December 2004}
\begin{abstract}
We present in this talk a recent analysis for the magnetic
moments of the baryon antidecuplet within the framework of the
chiral quark-soliton model with linear $m_s$ corrections 
considered.  We take into account the mixing of higher  
representations to the collective magnetic moment operator, which
comes from the SU(3) symmetry breaking.  Dynamical parameters of the
model are fixed by experimental data for the magnetic moments of the
baryon octet as well as by the masses of the octet, decuplet and of
$\Theta^{+}$.  The magnetic moment of $\Theta^{+}$ is rather
sensitive to the pion-nucleon sigma term and ranges from $-1.19\,{\rm n.m.}$
to $-0.33\,{\rm n.m.}$ as the sigma term is varied from $\Sigma_{\pi
  N} = 45$ to $75$ MeV, respectively.  On top of them, we obtained
that the strange magnetic moment of the nucleon has
the value of $\mu^{(s)}_N =+0.39$ n.m. within this scheme and turns
out to be almost independent of the sigma term.   
\end{abstract}
\maketitle
\section{Introduction}
Exotic pentaquark baryons has been a hot issue, since 
the LEPS collaboration~\cite{Nakano:2003qx} announced the new finding
of the $S=+1$ baryon $\Theta^+$ which was soon confirmed by 
a number of other experiments \cite{experiments}, together with an
observation of exotic $\Xi_{\overline {10}}$ states by the NA49
experiment at CERN~\cite{Alt:2003vb}, though it is still under
debate.  Those experiments searching for the pentaquark states was
stimulated by Diakonov et al.~\cite{Diakonov:1997mm}: Masses and decay 
widths of exotic baryon antidecuplet were predicted within the chiral
quark-soliton model.  

The discoveries of the pentaquark baryon $\Theta^{+}$ and possibly of
$\Xi_{\overline{10}}$ have triggered intensive theoretical
investigations (see, for example,
Refs.\cite{Jennings:2003wz,Zhu:2004xa,Goekeetal}).  The 
production mechanism of the $\Theta^{+}$ has been discussed in
Refs.\cite{Ohetal,Nametal,Yu:2003eq}. In particular, it is of great
interest to understand the photoproduction of the $\Theta^{+}$
theoretically, since the LEPS and CLAS collaborations used photons
as a probe to measure the $\Theta^{+}$. In order to describe the
mechanism of the pentaquark photoproduction, we have to know the
magnetic moment of the $\Theta^{+}$ and its strong coupling
constants. However, information on the static properties such as
antidecuplet magnetic moments and their strong coupling constants
is absent to date, so we need to estimate them theoretically.
Recently, two of the present authors calculated the magnetic
moments of the exotic pentaquarks, within the framework of the
chiral quark-soliton model~\cite{Kim:2003ay} in the chiral limit.
Since we were not able to fix all the parameters for the magnetic
moments in the chiral limit, we had to rely on the explicit model
calculations~\cite{Kim:1997ip,Kim:1998gt}.

A very recent work~\cite{Yang:2004jr} extended the analysis for the
magnetic moments of the baryons, taking into account the effect of
SU(3) symmetry breaking so that the necessary parameters are fixed by
the magnetic moments of the baryon octet.  In the present talk, we
would like to present the main results of Ref.~\cite{Yang:2004jr}.  

\section{Constraints on parameters}
\label{constraints}

The collective Hamiltonian describing baryons in the SU(3) chiral
quark-soliton model takes the following form \cite{Blotz:1992pw}:
\begin{equation}
\hat{H}=\mathcal{M}_{sol}+\frac{J(J+1)}{2I_{1}}+\frac{C_{2}(\text{SU(3)}%
)-J(J+1)-\frac{N_{c}^{2}}{12}}{2I_{2}}+\hat{H}^{\prime},
\label{Eq;1}
\end{equation}
with the symmetry breaking term given by:%
\begin{equation}
\hat{H}^{\prime}=\alpha D_{88}^{(8)}+\beta Y+\frac{\gamma}{\sqrt{3}}%
D_{8i}^{(8)}\hat{J}_{i},\label{Hsplit}%
\end{equation}
where parameters $\alpha$, $\beta$ and $\gamma$ are of order
$\mathcal{O}(m_{s})$ and are given as functions of the $\pi N$ sigma
term~\cite{Praszalowicz:2004dn}. $D_{ab}^{(\mathcal{R})}(R)$
denote SU(3) Wigner rotation matrices and $\hat{J}$ is a collective
spin operator.  The Hamiltonian given in Eq.(\ref{Hsplit}) acts on the
space of baryon wave functions: 
\begin{equation}
\left|  \mathcal{R}_{J},B,J_{3}\right\rangle=\sqrt{\mathrm{dim}%
(\mathcal{R})}(-1)^{J_{3}-Y^{\prime}/2}D_{Y,T,T_{3};Y^{\prime},J,-J_{3}%
}^{(\emph{R})\ast}(R).\label{Eq:wave_f}%
\end{equation}
Here, $\mathcal{R}$ stands for the allowed irreducible representations of the
SU(3) flavor group, \emph{i.e.} $\mathcal{R}=8,10,\overline{10},\cdots$ and
$Y,T,T_{3}$ are the corresponding hypercharge, isospin, and its third
component, respectively. Right hypercharge $Y^{\prime}=1$ is
constrained to be unity for the physical spin states for which $J$ and
$J_{3}$ are spin and its third component.  The {\em model-independent
approach} consists now in using Eqs.~(\ref{Eq;1}) and (\ref{Hsplit})
(and/or possibly analogous equations for other observables) and
determining model parameters such as $I_1,I_2,\alpha ,\beta, \gamma$
from experimental data.

The symmetry-breaking term (\ref{Hsplit}) of the collective
Hamiltonian mixes different SU(3) representations as follows:
\cite{Kim:1998gt}%
\begin{align}
\left|  B_{8}\right\rangle  &  =\left|  8_{1/2},B\right\rangle +c_{\overline
{10}}^{B}\left|  \overline{10}_{1/2},B\right\rangle +c_{27}^{B}\left|
27_{1/2},B\right\rangle ,\nonumber\\
\left|  B_{10}\right\rangle  &  =\left|  10_{3/2},B\right\rangle +a_{27}%
^{B}\left|  27_{3/2},B\right\rangle +a_{35}^{B}\left|  35_{3/2},B\right\rangle
,\nonumber\\
\left|  B_{\overline{10}}\right\rangle  &  =\left|  \overline{10}%
_{1/2},B\right\rangle +d_{8}^{B}\left|  8_{1/2},B\right\rangle +d_{27}%
^{B}\left|  27_{1/2},B\right\rangle +d_{\overline{35}}^{B}\left|
\overline{35}_{1/2},B\right\rangle, \label{admix}
\end{align}
where $\left|  B_{\mathcal{R}}\right\rangle $ denotes the state which reduces
to the SU(3) representation $\mathcal{R}$ in the formal limit
$m_{s}\rightarrow0$ and the spin index $J_{3}$ has been suppressed.
All relevant expressions for the mixing coefficients $c_{\mathcal R}^{B}$,
$a_{\mathcal R}^{B}$, and $d_{\mathcal R}^{B}$ can be found in  
Ref.~\cite{Yang:2004jr}.
\section{Magnetic moments in the chiral quark-soliton model}
\label{magmoms}

The collective operator for the magnetic moments can be parameterized
by six constants By definition in the {\em model-independent approach}
they are treated as free
~\cite{Kim:1997ip,Kim:1998gt}:%
\begin{eqnarray}
\hat{\mu}^{(0)} & =& w_{1}D_{Q3}^{(8)}\;+\;w_{2}d_{pq3}D_{Qp}^{(8)}
\cdot\hat{J}_{q}\;+\;\frac{w_{3}}{\sqrt{3}}D_{Q8}^{(8)}\hat{J}_{3},\cr
\hat{\mu}^{(1)} & =& \frac{w_{4}}{\sqrt{3}}d_{pq3}D_{Qp}^{(8)}D_{8q}
^{(8)}+w_{5}\left(
  D_{Q3}^{(8)}D_{88}^{(8)}+D_{Q8}^{(8)}D_{83}^{(8)}\right)\cr
&+& w_{6}\left(
  D_{Q3}^{(8)}D_{88}^{(8)}-D_{Q8}^{(8)}D_{83}^{(8)}\right). 
\end{eqnarray}
The parameters $w_{1,2,3}$ are of order $\mathcal{O}(m_{s}^0)$,
while $w_{4,5,6}$ are of order $\mathcal{O}(m_{s})$, $m_{s}$ being
regarded as a small parameter.

The full expression for the magnetic moments can be decomposed as follows:
\begin{equation}
\mu_{B}=\mu_{B}^{(0)}+\mu_{B}^{(op)}+\mu_{B}^{(wf)},
\end{equation}
where the $\mu_{B}^{(0)}$ is given by the matrix element of the
$\hat{\mu}^{(0)}$ between the purely symmetric states $\left|
\mathcal{R}_{J},B,J_{3}\right\rangle$, and the $\mu_{B}^{(op)}$ is
given as the matrix element of the $\hat{\mu}^{(1)}$ between the
symmetry states as well.  The wave function correction
$\mu_{B}^{(wf)}$ is given as a sum of the interference matrix elements
of the $\mu_{B}^{(0)}$ between purely symmetric states and admixtures
displayed in Eq.(\ref{admix}).  These matrix elements were calculated
for octet and decuplet baryons in Ref.\cite{Kim:1998gt}. 

The measurement of the $\Theta^{+}$ mass constrains the parameter
space of the model.  Recent phenomenological analyzes indicate that
our previous assumption on $\gamma$, i.e. $\gamma=0$, has to be most
likely abandoned. Therefore, our previous results for the magnetic 
moments of $8$, $10$ and $\overline{10}$ have to be reanalyzed. Now,
we show that a \emph{model-independent} analysis 
with this new phenomenological input yields $w_{2}$ much larger
than initially assumed, which causes $\mu_{\Theta^{+}}^{(0)}$ for
realistic values of $\Sigma_{\pi N}$ to be negative and rather
small.  Our previous results for the decuplet
magnetic moments turn out to hold within the accuracy of the model. 

The octet and decuplet magnetic moments were calculated in
Refs.\cite{Kim:1997ip,Kim:1998gt}.  For the antidecuplet
$\mu_{B}^{{\overline{10}}\;(0)}$ can be found in
Ref.\cite{Kim:2003ay}.  In order to calculate the $\mu_{B}^{(wf)}$, several
off-diagonal matrix elements of the $\hat{\mu}^{(0)}$ are
required. These have been calculated in
Ref.\cite{Praszalowicz:2004dn} in the context of the hadronic
decay widths of the baryon antidecuplet.

Denoting the set of the model parameters by
\begin{equation}
\vec{w}=(w_{1},\ldots,w_{6})
\end{equation}
the model formulae for the set of the magnetic moments in
representation 
$\mathcal{R}$ (of dimension $R$)%
\begin{equation}
\vec{\mu}^{\mathcal{R}}=(\mu_{B_{1}},\ldots,\mu_{B_{R}})
\end{equation}
can be conveniently cast into the form of the matrix equations:%
\begin{equation}
\vec{\mu}^{\mathcal{R}}=A^{\mathcal{R}}[\Sigma_{\pi N}]\cdot\vec{w},%
\end{equation}
where rectangular matrices $A^{8}$, $A^{10}$, and $A^{\overline{10}}$
can be found in Refs.\cite{Kim:1997ip,Kim:1998gt,Yang:2004jr}.  Note their
dependence on the 
pion-nucleon $\Sigma_{\pi N}$ term.  

\section{Results and discussion}\label{numres}
In order to find the set of parameters $w_{i}[\Sigma_{\pi N}]$, we
minimize the mean square deviation for the octet magnetic moments:
\begin{equation}
\Delta\mu^{8}=\frac{1}{7}\sqrt{\sum_{B}\left(
    \mu_{B,\,th}^{8}[\Sigma_{\pi N}]-\mu_{B,\,exp}^{8}\right)  ^{2}},
\end{equation}
where the sum extends over all octet magnetic moments, but the
$\Sigma^{0}$.  The value $\Delta\mu^{8}\simeq0.01$ is in practice
independent of the $\Sigma_{\pi N}$ in the physically interesting range
$45$ $-$ $75$ MeV.  The values of the $\mu_{B,\,th}^{8}[\Sigma_{\pi
N}]$ are independent of $\Sigma_{\pi N}$.   

Similarly, the value of the nucleon strange magnetic moment is independent of
$\Sigma_{\pi N}$ and reads $\mu_{N}^{(s)}=0.39 \,{\rm n.m.}$ in fair
agreement with our previous analysis of Ref.\cite{Kim:1998gt}.
Parameters $w_{i}$, however, do depend on $\Sigma_{\pi N}$.  This is
shown in Table.\ref{tablewi}:
\begin{table}[h]
\begin{tabular}
[c]{|c|cccccc|}\hline
\multicolumn{1}{|c|}{$\Sigma_{\pi N}$ [MeV]} & $w_{1}$ & $w_{2}$ &
$w_{3}$ &   $w_{4}$ & $w_{5}$ & $w_{6}$ \\ \hline
$45$ & $-8.564$ & $14.983$ & $7.574$ & $-10.024$ & $-3.742$ & $-2.443$\\
$60$ & $-10.174$ & $11.764$ & $7.574$ & $-9.359$ & $-3.742$ & $-2.443$\\
$75$ & $-11.783$ & $8.545$ & $7.574$ & $-6.440$ &  $-3.742$ & $-2.443$
\\ \hline
\end{tabular}
\caption{Dependence of the parameters $w_i$ on $\Sigma_{\pi N}$.}
\label{tablewi}
\end{table}
Note that parameters $w_{2,3}$ are formally
$\mathcal{O}(1/N_{c})$ with respect to $w_{1}$. For smaller
$\Sigma_{\pi N}$, this $N_{c}$ counting is not borne by explicit
fits.  The $\mu_B^{(0)}$ can be parametrized by the following
two parameters $v$ and $w$:
\begin{equation}%
\begin{array}
[c]{ccrcc}%
v & = & \left(  2\mu_{\mathrm{n}}-\mu_{\mathrm{p}}+3\mu_{\Xi^{0}}+\mu_{\Xi
^{-}}-2\mu_{\Sigma^{-}}-3\mu_{\Sigma^{+}}\right)  /60 & = & -0.268,\\
w & = & \left(  3\mu_{\mathrm{p}}+4\mu_{\mathrm{n}}+\mu_{\Xi^{0}}-3\mu
_{\Xi^{-}}-4\mu_{\Sigma^{-}}-\mu_{\Sigma^{+}}\right)  /60 & = & 0.060.
\end{array}
\label{Eq:mean}%
\end{equation}
which are free of linear $m_{s}$ corrections~\cite{Kim:1998gt}.
This is a remarkable feature of the present fit, since when the
$m_{s}$ corrections are included, the $m_{s}$-independent parameters
need not be refitted.  This property will be used in the following
when we restore the linear dependence of the $\mu_{B}^{\overline{10}}$
on $m_{s}$. 

The magnetic moments of the baryon decuplet and
antidecuplet depend on the $\Sigma_{\pi N}$.  However, the
dependence of the decuplet is very weak, which as summarized
in Table \ref{table10},
\begin{table}[h]
\begin{tabular}
[c]{|c|cccc|}\hline
$\Sigma_{\pi N}$ [MeV] & $\Delta^{++}$ & $\Delta^{+}$ & $\Delta^{0}$ & $\Delta^{-}$ \\ \hline
$45$ & $5.40$ & $2.65$ & $-0.09$ & $-2.83$ \\
$60$ & $5.39$ & $2.66$ & $-0.08$ & $-2.82$ \\
$75$ & $5.39$ & $2.66$ & $-0.07$ & $-2.80$ \\
\text{Ref.\cite{Kim:1997ip}} & $5.34$ & $2.67$ & $-0.01$ & $-2.68$ \\ \hline
\end{tabular}
\begin{tabular}
[c]{|c|cccccc|}\hline
$\Sigma_{\pi N}$ [MeV] & $\Sigma^{\ast+}$ & $\Sigma^{\ast0}$ &
$\Sigma^{\ast-}$ & $\Xi^{\ast0}$ & $\Xi^{\ast-}$ &
$\Omega^{-}$\\\hline 
$45$ & $2.82$ & $0.13$ & $-2.57$& $0.34$ & $-2.31$ & $-2.05$\\
$60$ & $2.82$ & $0.13$ & $-2.56$& $0.34$ & $-2.30$ & $-2.05$\\
$75$ & $2.81$ & $0.13$ & $-2.55$& $0.33$ & $-2.30$ & $-2.05$\\
\text{Ref.\cite{Kim:1997ip}} & $3.10$ & $0.32$ & $-2.47$ & $0.64$ &
$-2.25$ & $-2.04$ \\ \hline 
\end{tabular}
\caption{Magnetic moments of the baryon decuplet.}
\label{table10}%
\end{table}
where we also display the theoretical predictions from
Ref.\cite{{Kim:1997ip}} for $p=0.25$.  Let us note that the $m_{s}$
corrections are not large for the decuplet and the approximate
proportionality of the $\mu_{B}^{10}$ to the baryon charge
$Q_{B}$ still holds.

\begin{figure}[h]
\begin{center}
\includegraphics[scale=1.1]{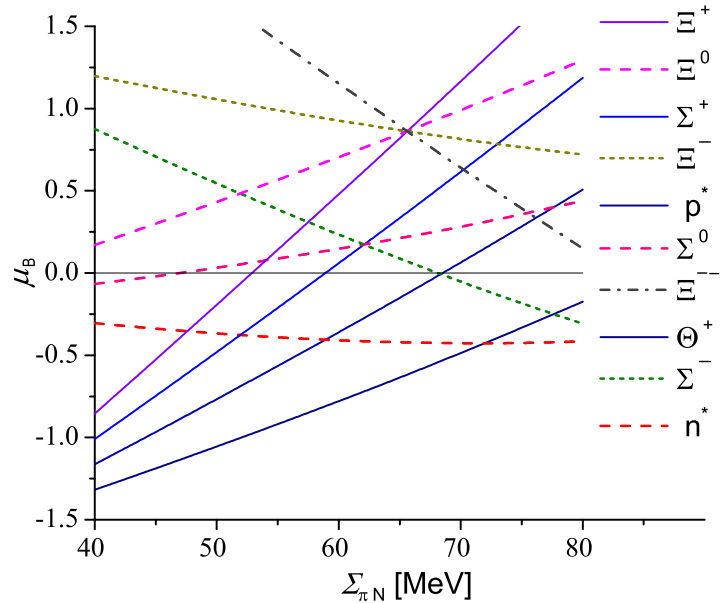}
\end{center}
\caption{Magnetic moments of antidecuplet as functions of $\Sigma_{\pi N}$.}%
\label{fig:10bar}%
\end{figure}

Finally,  for antidecuplet we have a strong dependence on
$\Sigma_{\pi N}$, yielding the numbers of Table \ref{tab10b}.
\begin{table}[h]
\begin{tabular}
[c]{|c|cccccc|}\hline
$\Sigma_{\pi N}$ [MeV] & $\Theta^{+}$ & $p^{\ast}$ & $n^{\ast}$
&$\Sigma_{\overline{10}}^{+}$ & $\Sigma_{\overline{10}}^{0}$ &
$\Sigma_{\overline{10}}^{-}$ \\ \hline 
$45$ & $-1.19$ & $-0.97$ & $-0.34$ & $-0.75$ & $-0.02$ & $\;0.71$ \\
$60$ & $-0.78$ & $-0.36$ & $-0.41$ & $\;0.06$ & $\;0.15$ & $\;0.23$ \\
$75$ & $-0.33$ & $\;0.28$ & $-0.43$ & $\;0.90$ & $\;0.36$ & $-0.19$ \\
\hline 
\end{tabular}
\begin{tabular}
[c]{|c|cccc|}\hline
$\Sigma_{\pi N}$ [MeV] & $\Xi_{\overline{10}}^{+}$ &
$\Xi_{\overline{10}}^{0}$ & $\Xi_{\overline{10}}^{-}$ &
$\Xi_{\overline {10}}^{--}$ \\\hline
$45$ & $-0.53$ & $0.30$ & $1.13$ & $1.95$\\
$60$ & $\;0.48$ & $0.70$ & $0.93$ & $1.15$\\
$75$ & $\;1.51$ & $1.14$ & $0.77$ & $0.39$ \\ \hline
\end{tabular}
\caption{Magnetic moments of the baryon antidecuplet.}
\label{tab10b}%
\end{table}
The results listed in Table~\ref{tab10b} are further depicted in
Fig.\ref{fig:10bar}.

The wave function
corrections cancel for the non-exotic baryons and add
constructively for the baryon antidecuplet.  In particular, for
$\Sigma_{\pi N}=75$~MeV we have large admixture coefficient of
27-plet: $d_{27}^{B}$ tends to dominate otherwise small magnetic
moments of antidecuplet.  At this point, the reliability of the
perturbative expansion for the antidecuplet magnetic moments may
be questioned.  On the other hand, as remarked above, the $N_{c}$
counting for the $w_{i}$ coefficients works much better for large
$\Sigma_{\pi N}$.  One notices for reasonable values of
$\Sigma_{\pi N}$ some interesting facts, which were partially
reported already in Ref.\cite{Kim:2003ay}: The magnetic moments of
the antidecuplet baryons are rather small in absolute value. For
$\Theta^+$ and $p^*$ one obtains negative values although the
charges are positive.  For $\Xi^{-}_{\overline{10}}$ and
$\Xi^{--}_{\overline{10}}$ one obtains positive values although
the signs of the charges are negative.

\section{Conclusion and summary}\label{summary}

Our present analysis shows that $\mu_{\Theta^+}< 0$, although the
magnitude depends strongly on the model parameters.  The measurement
of $\mu_{\Theta^+}$ could therefore discriminate between different
models. This also may add to reduce the ambiguities in the
pion-nucleon sigma term $\Sigma_{\pi N}$.

In the present work, we determined the magnetic moments of the
baryon antidecuplet in the \emph{model-independent} analysis
within the chiral quark-soliton model, i.e. using the rigid-rotor
quantization with the linear $m_{s}$ corrections included.
Starting from the collective operators with dynamical parameters
fixed by experimental data, we obtained the magnetic moments of
the baryon antidecuplet.  The expression for the magnetic moments of
the baryon antidecuplet is different from those of the baryon
decuplet. We found that the magnetic moment $\mu_{\Theta^{+}}$ is
negative and rather strongly dependent on the value of the
$\Sigma_{\pi N}$.  Indeed, the $\mu _{\Theta^{+}}$ ranges from
$-1.19\,{\rm n.m.}$ to $-0.33\,{\rm n.m.}$ for $\Sigma_{\pi N} = 45$
and $75$ MeV, respectively.  

\section*{Acknowledgments}
H.-Ch.K and G.-S. Y are very much thankful to the
organizers of the Workshop Pentaquark 04, in particular, T. Nakano and
A. Hosaka for their hospitality.  H.-Ch.K is grateful to J.K. Ahn,
S.I. Nam, M.V. Polyakov, and I.K. Yoo for valuable discussions.  The
present work is supported by Korea Research Foundation Grant:
KRF-2003-041-C20067 (H.-Ch.K.) and by the Polish State Committee for
Scientific Research under grant 2 P03B 043 24 (M.P.) and by
Korean-German and Polish-German grants of the Deutsche
Forschungsgemeinschaft.

\end{document}